\documentclass[%
 aps,
 prb,
 amsmath,
 amssymb,
preprint,
]{revtex4-1}
\usepackage[dvipdfmx]{graphicx}
\usepackage{dcolumn}
\usepackage{braket}
\usepackage[mathlines]{lineno}
\usepackage{bm}
\usepackage{color}
\usepackage{ulem}

\usepackage{enumerate}
\usepackage{algorithm, algorithmic}

\draft 
\graphicspath{{./}}

\usepackage{tabularx}
\newcolumntype{C}{>{\centering\arraybackslash}X}
\newcolumntype{L}{>{\raggedright\arraybackslash}X}
\newcolumntype{R}{>{\raggedleft\arraybackslash}X}
\usepackage {threeparttable}

\begin{document}
\title{Contour integral method for obtaining the self-energy matrices of electrodes in electron transport calculations}

\author{Shigeru Iwase}
\email{iwase@ccs.tsukuba.ac.jp}
\affiliation{Department of Physics, University of Tsukuba, Tsukuba, Ibaraki, 305-8571, Japan}
\author{Yasunori Futamura}
\affiliation{Department of Computer Science, University of Tsukuba, Tsukuba, Ibaraki, 305-8573}
\author{Akira Imakura}
\affiliation{Department of Computer Science, University of Tsukuba, Tsukuba, Ibaraki, 305-8573}
\author{Tetsuya Sakurai}
\affiliation{Department of Computer Science, University of Tsukuba, Tsukuba, Ibaraki, 305-8573}
\author{Shigeru Tsukamoto}
\address{Peter Gr{\"u}nberg Institut \& Institute for Advanced Simulation, Forschungszentrum J{\"u}lich and JARA, D-52425 J{\"u}lich, Germany}
\author{Tomoya Ono}
\affiliation{Department of Physics, University of Tsukuba, Tsukuba, Ibaraki, 305-8571, Japan}
\affiliation{Center for Computational Sciences, University of Tsukuba, Tsukuba, Ibaraki, 305-8577, Japan}
\date{\today}

\begin{abstract}
We propose an efficient computational method for evaluating the self-energy matrices of electrodes to study ballistic electron transport properties in nanoscale systems. To reduce the high computational cost incurred in large systems, a contour integral eigensolver based on the Sakurai-Sugiura method combined with the shifted biconjugate gradient method is developed to solve exponential-type eigenvalue problem for complex wave vectors. A remarkable feature of the proposed algorithm is that the numerical procedure is very similar to that of conventional band structure calculations. We implement the developed method in the framework of the real-space higher-order finite difference scheme with nonlocal pseudopotentials. Numerical tests for a wide variety of materials validate the robustness, accuracy, and efficiency of the proposed method. As an illustration of the method, we present the electron transport property of the free-standing silicene with the line defect originating from the reversed buckled phases.
\end{abstract}

\maketitle

\section{Introduction}
First-principles simulations based on density functional theory\cite{Hohenberg1964,Kohn1965} (DFT) and the Landauer-B\"{u}ttiker formalism \cite{Landauer1957,Buttiker1985,Datta1995} are recognized as a powerful tool for understanding and predicting the electron transport properties of nanoscale systems. Examples of such computational methods include the Lippmann-Schwinger scattering approach,\cite{Lang1995,Ventra2000,Ventra2001} recursive transfer-matrix method,\cite{Hirose1994,Hirose1995} wave function matching (WFM) method,\cite{Choi1999,Fujimoto2003} and nonequilibrium Green's function (NEGF) method.\cite{Tayer2001,Brandbyge2002} Among these, the NEGF method combined with the localized basis has been used extensively in the field of molecular electronics and spintronics, because Green's function is compactly described by the localized basis and bound states can be easily included by the contour integral of the retarded Green's function. However, the transport properties obtained using the localized basis sometimes significantly depend on the size of the basis set,\cite{Ke2007,Strange2008,Reuter2013} and they need to be checked by comparing the results using a plane-wave basis or real-space grid whose convergence with respect to the basis set size is straightforward. Furthermore, it is well known that the electron transport in the tunneling region is difficult to address with the localized basis set approach due to the incompleteness of the basis sets. This tunneling problem is handled by putting the ghost orbital in the vacuum region or increasing the cutoff radius of the basis set. However, the problem with accuracy and efficiency persists.\cite{Garcia-Lekue2010,Zotti2017} Electron transport calculations using a plane-wave basis exist.\cite{Choi1999,Wang2005,Wang2006} While the plane-wave basis method enables highly accurate simulations, it requires that left and right electrodes to be identical, because of the artificial periodic boundary condition associated with the plane-wave basis, which leads to a strong limitation on the computational model of transport calculations. In addition, the size of the Hamiltonian matrix that is represented in the plane-wave basis is much larger than that in the localized basis, and it is difficult to parallelize; this hampers large-scale electron transport calculations. On the other hand, real-space grid approach is free from such drawbacks and its Hamiltonian matrix is highly sparse; this not only facilitates implementation on parallel computers but also enables fast computation using sophisticated numerical algorithms designed for sparse matrices. 

Several real-space grid methods have been developed over the last decade. In 2003, Fujimoto and Hirose developed the overbridging boundary matching (OBM) method \cite{Fujimoto2003} for calculating the generalized Bloch states of electrodes and the scattering states, and they applied it to a gold atomic wire. Subsequently, Khomyakov  $et~al$. \cite{Khomyakov2004} proposed a real-space grid implementation of the WFM method formulated by Ando.\cite{Ando1994} However, the applications of these methods, including extensions of the OBM method,\cite{Ono2012,Tsukamoto2014,Iwase2015} have remained limited to relatively small systems or very rough approximations (e.g., first-order finite-difference, local pseudopotential, and jellium electrode) owing to the large computational cost of inverting Hamiltonian matrices. As an extension of the OBM method, Kong $et~al$. \cite{Kong2006,Kong2007} developed a method for computing the scattering states without explicitly inverting the Hamiltonian in the transition region, as shown in Fig.~\ref{fig:transport-model}; hereafter, we refer to their method as the improved OBM (IOBM) method. Nevertheless, the IOBM method still requires the inversion of the Hamiltonian matrix in electrode regions and the computation for solving the generalized eigenvalue problem with very dense matrices. To avoid these inversion-related problems, Feldman $et~al$. \cite{Feldman2014,Feldman2016} recently proposed a transport calculation method based on Green's function by using the adsorbing boundary condition instead of the self-energy matrices of electrodes. However, the use of an adsorbing boundary condition requires several parameters to be tuned manually to remove spurious reflections at the boundaries; this may restrict its applicability to complicated electrode materials. In addition, their approach can only be used for the linear response scheme, that is, for non-self-consistent calculations. Overall, compared with mature conventional band structure calculations for a real-space grid approach,\cite{chelikowsky1994prl,chelikowsky1994prb,ICP} currently available methods for real-space transport calculations are computationally too expensive for investigating the transport properties of nanoscale systems within a realistic timeframe. In particular, a reliable and practical method is not available for evaluating the self-energy matrices of electrodes using higher-order finite-differences and nonlocal pseudopotentials.

In this article, we present a method for handling the most computationally expensive part of real-space transport calculations, namely, the computation of self-energy matrices. Because the proposed method shows significantly improved execution time and excellent parallel efficiency compared with methods involving an explicit matrix inversion and a large, dense eigenvalue problem, we can readily perform challenging real-space transport calculations that were not feasible thus far. The proposed method involves the following procedures: express the Kohn-Sham (KS) equation as an exponential-type eigenvalue problem (EEP) for complex wave vectors $k$, compute only those solutions that dominantly contribute to electron transport, and construct self-energy matrices by using them. To solve EEP efficiently, we developed a contour integral method based on the Sakurai-Sugiura (SS) method \cite{Sakurai2003,Asakura2009} combined with the shifted biconjugate gradient (BiCG) method.\cite{Frommer2003} A remarkable feature of the SS method is that it replaces the difficulty of solving the nonlinear eigenvalue problem with the difficulty of solving linear systems with multiple right-hand sides:
\begin{equation}
\label{sle}
[E-H(k)]Y=V,
\end{equation}
where $E$, $V$, and $H(k)$ are the input energy, random vectors, and Hamiltonian matrix with a complex wave vector $k$, respectively (see Sec.~\ref{sec:subsec3.2} for details). Because $H(k)$ has the same matrix structure as the Hamiltonian matrix used in the conventional band structure calculation, we can use the advanced numerical techniques for the conventional band calculation to solve the above equation. Therefore, this method is fast, and it overcomes all limitations of previous real-space grid methods, such as large computational cost and memory consumption associated with the matrix inversion, and the treatment of higher-order finite differences and nonlocal pseudopotentials, while keeping the relatively high accuracy. In addition, as discussed in Sec.~\ref{sec:subsec3.1}, this method provides some general advantages compared to previous similar approaches \cite{Sorensen2008,Laux2012,Iwase2017,Bruck2017} under harsh conditions that other approaches cannot handle easily; we believe that such advantages will be essential to the widespread use of the proposed method. Finally, it should be noted that the proposed method is highly versatile and that it can be implemented in various schemes such as localized basis sets, together with the use of the (sparse) LU decomposition for solving Eq.~\eqref{sle}.

The remainder of the paper is organized as follows. Sec.~\ref{sec:sec2} briefly summarizes the basic formulation of the WFM method that is necessary to understand our implementation. In Sec.~\ref{sec:sec3}, we present an efficient implementation for evaluating the self-energy matrices. In Sec.~\ref{sec:sec4}, we demonstrate the accuracy, robustness, and efficiency of the proposed method through a series of test calculations. Sec.~\ref{sec:sec5} presents transport calculations using the self-energy matrices computed with the proposed method. In Sec.~\ref{sec:Application}, a first-principles analysis of electron transport properties of the free-standing silicene with the line defect originating from the reverse buckled phases is presented to illustrate the proposed method. Finally, Sec.~\ref{sec:sec6} presents the conclusions of this study.

\label{sec:1}

\section{General formalism}
\label{sec:sec2}
\begin{figure*}[t]
\begin{center}
\includegraphics[width=120mm]{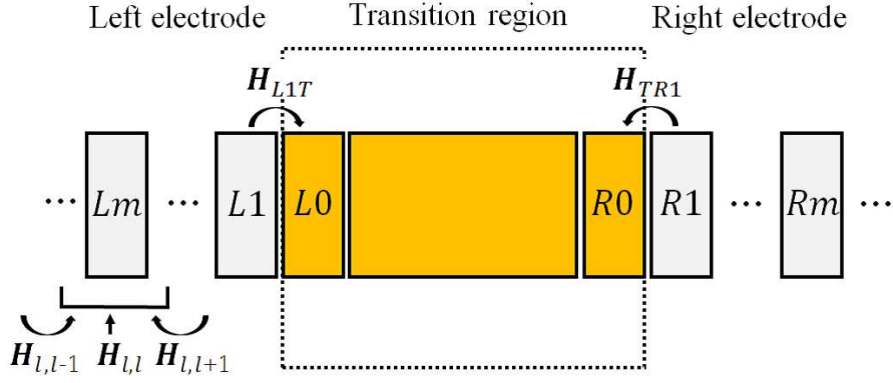}
\caption{Schematic representation of quasi-one-dimensional conductor sandwiched between two semi-infinite electrodes. Because of the localized feature of the real-space grid approach, semi-infinite crystalline electrodes can be divided into principal layers denoted by $L0,L1,\cdots,Lm;R0,R1,\cdots,Rm$. $L0$ and $R0$ are incorporated into the transition region as WFM planes.}
\label{fig:transport-model}
\end{center}
\end{figure*}

The implementation of the WFM method with a real-space grid approach using higher-order finite differences has been reported in detail in Ref.~\onlinecite{ICP}. Herein, we briefly review details that are essential to understanding our method for evaluating the self-energy matrices of electrodes. We consider a quasi-one-dimensional conductor sandwiched by two semi-infinite electrodes, as shown in Fig.~\ref{fig:transport-model}. Owing to the localized feature of the KS Hamiltonian in the real-space grid approach, semi-infinite crystalline electrodes can be divided into the principal layers $(L0,L1,\cdots,Lm;R0,R1,\cdots,Rm)$; these are defined as the smallest layers such that only nearest-neighbor interactions exist between principal layers. Typically, the size of the principal layer is much smaller than that of the unit cell of the electrode. As in Ref.~\onlinecite{Ono2012}, the transition region includes at least two principal layers $L0$ and $R0$ as matching planes of the wave function. To avoid complex notations, we assume that the left and right electrodes are identical. The KS equation for the entire region is given as follows:
\begin{equation}
\label{KS}
H \ket{\psi} = E \ket{\psi},
\end{equation}
where $H$ is the KS Hamiltonian of the entire system and $\ket{\psi}$ is the scattering state. For simplicity, we restrict our formulation to the real-space norm-conserving pseudopotential method. However, extensions to the ultrasoft pseudopotential \cite{Vanderbilt1990} and projector augmented wave method \cite{Blochl1994} are possible.

\subsection{Generalized Bloch states}
\label{sec:subsec2.1}

Generalized Bloch states (i.e., Bloch states with a complex wave vector) play a significant role in the WFM method, in which the scattering state is expanded by them at the left and right matching planes. To find the generalized Bloch states, we seek the complex wave vectors that satisfy Eq.~\eqref{KS} under a periodic boundary condition. Because of the local structure of the KS Hamiltonian in a real-space grid approach, it suffices to consider the coupling between the nearest-neighbor unit cells of the electrode; then, Eq.~\eqref{KS} can be written as the following matrix equation:
\begin{equation}
\label{KS_R}
-{H}_{l,l-1}\psi_{l-1}+(E-{H}_{l,l})\psi_{l}-{H}_{l,l+1}\psi_{l+1}=0,
\end{equation}
where $\psi_{l}$ is the $N$-dimensional vector in the $l$-th unit cell and ${H}_{i,j}$ is the $N \times N$ KS Hamiltonian matrix between the $i$-th and $j$-th unit cell, with $N$ being the total number of real-space grids in the unit cell of the electrode. By introducing the Bloch ansatz, the above equation can be rewritten as
\begin{equation}
\label{eq:QEP}
[-e^{-{\rm i}k_na}H_{l,l-1}+(E-H_{l,l})-e^{{\rm i}k_na} H_{l,l+1}]\phi_{n}=0,
\end{equation}
with $n$-th complex wave vector $k_n$ and $a$ is the distance between adjacent unit cells. $\{k_n\}$ and $\{\phi_n\}$ are to be determined by solving Eq.~\eqref{eq:QEP} for a given input energy $E$. It is worth noting that the coupling matrices $H_{l,l-1}$ and $H_{l,l+1}$ are extremely sparse, and many columns are all zero. Because of the singularity of the coupling matrices, solutions of Eq.~\eqref{eq:QEP} will have Im$(k_n) \rightarrow \pm \infty$. However, the contribution of the waves that decay infinitely fast is negligibly small, and thus it is sufficient to compute $2r$ physically important solutions, with $r$ being the rank number of the coupling matrix computed with the singular value decomposition technique.\cite{Rungger2008, Tsukamoto2014,Reuter2016}
 Furthermore, $2r$ solutions can be classified into $r$ left-going waves $\{\phi_n^{-}\}$ with either Im$(k_n)<0$ or Im$(k_n)=0$ and $\partial E / \partial k_n < 0$ and $r$ right-going waves $\{ \phi_n^{+}\}$ with either Im$(k_n)>0$ or Im$(k_n)=0$ and $\partial E / \partial k_n > 0$. This is because if $k_n$ is an eigenvalue, then $-k_n$, $k_n^{\ast}$, and $-k_n^{\ast}$ are also eigenvalues.\cite{Boykin1996}

\subsection{Construction of self-energy matrices from generalized Bloch states}
\label{sec:subsec2.2}

Next, we define the self-energy matrices of electrodes by using the generalized Bloch states. As in Ref.~\onlinecite{Khomyakov2005}, we introduce $M$-dimensional dual vectors $\widetilde \phi^{-}_{Lm,i}$ of the left-going waves $\phi^{-}_{Lm,i}$ in the $m$-th principal layers of the left electrode to satisfy $(\widetilde \phi^{-}_{Lm,i})^{\dagger} \phi^{-}_{Lm,j}=\delta_{i,j}$ as well as $\widetilde \phi^{+}_{Rm,i}$ of the right-going waves $\phi^{+}_{Rm,i}$ in the $m$-th principal layers of the right electrode to satisfy $(\widetilde \phi^{+}_{Rm,i})^{\dagger} \phi^{+}_{Rm,j}=\delta_{i,j}$, with $M$ being the total number of real-space grid points in the principal layer. When we also define the $M \times r$ matrices ${Q}_{Lm}^{-},{Q}_{Rm}^{+},\widetilde {Q}_{Lm}^{-}$, and $\widetilde {Q}_{Rm}^{+}$ as
\begin{eqnarray}
{Q}_{Lm}^{-}&=&(\phi^{-}_{Lm,k_1},\phi^{-}_{Lm,k_2},...,\phi^{-}_{Lm,k_r}), \label{eq:Bloch_mat1}\\
{Q}_{Rm}^{+}&=&(\phi^{+}_{Rm,k_1},\phi^{+}_{Rm,k_2},...,\phi^{+}_{Rm,k_r}), \\
\widetilde {Q}_{Lm}^{-}&=&(\widetilde \phi^{-}_{Lm,k_1}, \widetilde \phi^{-}_{Lm,k_2},...,\widetilde \phi^{-}_{Lm,k_r}), \\
\widetilde {Q}_{Rm}^{+}&=&(\widetilde \phi^{+}_{Rm,k_1},\widetilde \phi^{+}_{Rm,k_2},...,\widetilde \phi^{+}_{Rm,k_r}), \label{eq:Bloch_mat4}
\end{eqnarray}
$M \times M$ self-energy matrices of the left and right electrodes, $\Sigma_{L0}$ and $\Sigma_{R0}$ can be defined via the WFM formulation \cite{ICP,Khomyakov2005,Ono2012} as
\begin{equation}
\label{eq:SE}
{\Sigma}_{L0}\!=\!{H}_{L1T}{Q}_{L1}^{-}(\widetilde {Q}_{L0}^{-})^{\dagger}, \quad {\Sigma}_{R0}\!=\!{H}_{TR1}{Q}_{R1}^{+}(\widetilde {Q}_{R0}^{+})^{\dagger},
\end{equation}
where ${H}_{L1T}$ (${H}_{TR1}$) is the $M \times M$ KS Hamiltonian matrix between the transition region and $L_1$ $(R_1)$ region. In Eqs.~\eqref{eq:Bloch_mat1}-\eqref{eq:SE}, the labels “L” and “-” (“R” and “+”) always appear pairwise because generalized Bloch states with Im$(k_n)\neq0$ must vanish deep inside the left (right) electrode.

It should be noted that the self-energy matrices defined by Eq.~\eqref{eq:SE} are equivalent to those defined by the surface Green's function.\cite{Sanvito1999,Rungger2008} The iterative technique is used most commonly to evaluate the surface Green's function with the localized basis.\cite{Sancho1984,Sancho1985} However, this technique is virtually impossible to apply to the real-space grid approach because it requires the matrix inversion of very dense matrices with the dimension of real-space grids in the unit cell, which is typically over 100000.

\subsection{Decaying behavior of generalized Bloch states}
\label{sec:subsec2.3}

\begin{table*}
\centering
\caption{Number of right-going waves that satisfy $10^{-8} \le |e^{{\rm i}k_na}|^{l} \le 1$ for several electrode materials as a function of the number of unit cells $l$. The Fermi energy is used as an input energy, and all calculations are performed using the OBM method. Geometry descriptions are summarized in Table~\ref{table:runtime}.}
{
\label{table:details_calc}
\begin{tabularx}{170mm}{CCCCCCC}
\hline \hline
Material & \#solutions & $l=1$ & $l=2$ &$l=3$ &$l=4$ & $l\rightarrow \infty$\\
\hline
Au chain & 21632 &  102 &  17 & 6  & 4 & 1 \\
Al(100) wire & 51200 & 334 & 83 & 38 & 20 & 7 \\
(6,6)CNT & 41472 &  997 & 278 & 126 & 77 & 2 \\
Graphene & 7168 & 99 & 32 & 14 & 11 & 2 \\
Silicene   & 24576  &   64      &  19     &   13     &   11   & 2 \\
Au(111) bulk & 7680 & 63 & 40 & 24 & 20 & 3 \\
\hline \hline
\end{tabularx}
}
\centering
\end{table*}

The generalized Bloch states include both propagating and decaying waves. It is instructive to note that a large majority of the generalized Bloch waves in a real-space grid decay completely within one unit cell of the electrode, and therefore, they contribute little to the electron transport. Table~\ref{table:details_calc} shows the number of right-going waves that satisfy $10^{-8} \le |e^{{\rm i}k_na}|^{l} \le 1$ for several electrode materials as a function of the number of unit cells $l$. We use the Fermi energy as an input energy. In all systems, the number of right-going waves satisfying the above criterion is relatively small compared to the total number of nontrivial solutions even when $l=1$. From the Bloch ansatz, it is clear that the majority of right-going waves decay so fast that they contribute negligibly to the electron transport if we add 1--2 unit cells as a buffer layer. The left-going waves behave in exactly the same manner as the right-going waves because their eigenvalues are pairwise.

\begin{figure*}[t]
\begin{center}
\includegraphics[width=120mm]{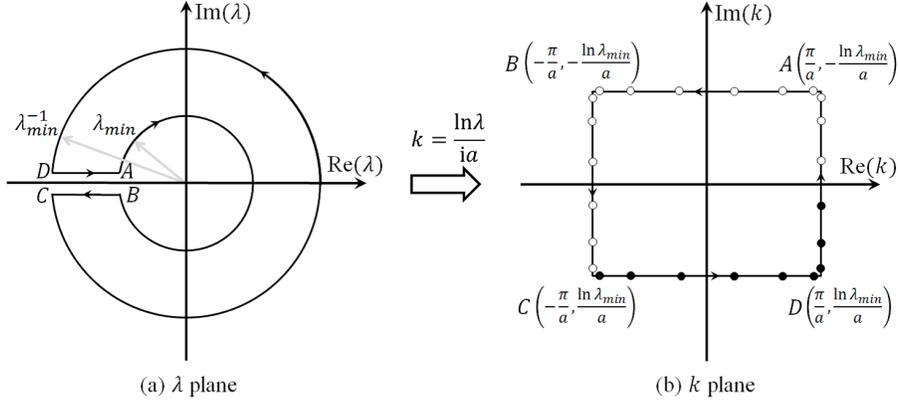}
\caption{Two equivalent contours in the complex $\lambda$ plane and $k$ plane. (a) A multiply connected domain between two concentric circles at the origin of radii $\lambda_{min}$ and $\lambda_{min}^{-1}$ in the $\lambda$ plane. (b) A simply connected domain obtained from (a) by variable conversion, $k=\ln{\lambda}/{\rm i}a$. The contour integral proceeds in a counter-clockwise direction along a rectangular side path. For this type of contour, quadrature points are specified by a pair of polynomial orders of the Gauss-Legendre quadrature rule, that is, $N_q=(N_{q1},N_{q2})$. In the figure, there are $N_{q1}=6$ Gauss-Legendre quadrature points along the Re$(k)$ axis and $N_{q2}=6$ points along the Im$(k)$ axis. The total number of quadrature points is $N_{int}=2N_{q1}+2N_{q2}$. The quadrature points are indicated by $\circ$ and $\bullet$; however, the numerical integration can only be performed using the values indicated by $\bullet$ owing to the symmetry of the Hamiltonian matrix $H(k)$.}
\label{fig:ring}
\end{center}
\end{figure*}

Consequently, the self-energy matrices can be well approximated by a relatively small number of propagating and moderately decaying waves that correspond to the solutions of Eq.~\eqref{eq:QEP} with $\lambda_n(=e^{{\rm i}k_na})$ being close to the unit circle in the complex plane, that is,
\begin{equation}
\lambda_{min} \le |\lambda_n| \le \lambda_{min}^{-1},
\label{eq:lambda_cut}
\end{equation}
where $\lambda_{min}$ is the radius of the inner circle in Fig.~\ref{fig:ring}(a). If $\lambda_{min}$ is set to a reasonably small value, the 
transport calculations remain accurate.\cite{Khomyakov2004,Sorensen2008} Actually, as shown in the test section, the results obtained using the approximated self-energy matrices that remove fast-decaying waves are visibly indistinguishable from those obtained using exact ones, and they reproduce the previous plane-wave transport calculations accurately. Therefore, we do not distinguish between approximated and exact self-energy matrices except when comparing both computational results.

\section{Implementation}
\label{sec:sec3}

The computation of self-energy matrices is one of the bottlenecks in electron transport calculations, especially in the real-space grid approach. If we follow the WFM method, the most computationally demanding part is determining the generalized Bloch states, that is, solving Eq.~\eqref{eq:QEP} for a given energy point. In this section, we first discuss some advantages of the proposed method compared to previous methods based on the same strategy, and then, we present the algorithmic details of this method.

\subsection{Variable conversion from $\lambda$ space to $k$ space}
\label{sec:subsec3.1}
Several efficient approaches have been proposed to solve Eq.~\eqref{eq:QEP} as a quadratic eigenvalue problem (QEP) for $\lambda$ and to determine the eigenvalues inside the contour shown in Fig.~\ref{fig:ring}(a). From the viewpoint of the eigensolver, these approaches are classified into two types: (i) shift-and-invert Krylov subspace approach\cite{Sorensen2008} and (ii) contour integral approach using contour integration based on the SS method,\cite{Iwase2017} Polizzi's FEAST method,\cite{Laux2012} and Beyn method.\cite{Bruck2017} All these approaches have been demonstrated to work successfully when $0.1 < \lambda_{min} < 1$; however, they are inefficient and unstable when $\lambda_{min} \ll 0.1$. For example, the shift-and-invert Krylov subspace approach is designed for determining eigenvalues close to a given shift, and thus, it is unsuitable for searching all target eigenvalues distributed in a wide range of the complex plane. On the other hand, in the contour integral approach, the number of projectors has to be increased when the contour is enlarged. For our target problem, the BiCG method is employed as a solver for linear systems to obtain the projectors, indicating that the computational cost linearly scales as the number of right-hand sides. The current authors' group demonstrated that the moment-based contour integral approach is appropriate when the linear systems are solved by the CG method because the number of right-hand sides is successfully reduced by introducing the moment.\cite{Futamura2013} In this subsection, we focus on some numerical difficulties faced in previous approaches based on the moment-based contour integral approach.\cite{Iwase2017}

In general, the moment-based contour integral method generates a subspace spanned by eigencomponents inside a given region and extracts target eigenpairs from this subspace. However, if $\lambda_{min} \ll 0.1$, the subspace quality degrades owing to two reasons. First, the large difference in radius between the inner and the outer circles causes a significant round-off error because the contour integrations along the inner and outer circle give small and large values, respectively. Consequently, the information obtained from the contour integrations along the inner circle will not be properly contained, and significant accuracy degradation will occur from the outer circle toward the inner circle (see Sec.~\ref{sec:subsec4.2}). Although this round-off error can be suppressed by introducing a number of circles between the inner and outer circles, the additional computational cost incurred owing to the added circles makes the algorithm inefficient. Second, unneeded eigencomponents contaminate the subspace. In the actual calculation, because the contour integration is performed numerically, the subspace involves eigencomponents both inside the ring-shaped region and in the vicinity of the circles. Because of the pairwise relationship between the inner and the outer eigenvalues $(\lambda_n,\lambda_n^{*-1})$, too many eigenvalues exist in the vicinity of the inner circle, and thus, the subspace size increases. This incurs a large computational cost and deteriorates the accuracy of the target eigenpairs. Moreover, the efficient implementation introduced in the later subsection cannot be used.

To avoid the numerical difficulties arising from the explicit computation of eigenvalues in the $\lambda$ plane, we convert the variable space from the $\lambda$ plane to the $k$ plane, as shown in Fig.~\ref{fig:ring}. Through the variable conversion $k=\ln \lambda/{\rm i}a$, the ring-shaped region is replaced with the rectangular region, as shown in Fig.~\ref{fig:ring}(b). The contour integration along the rectangular region can be performed without suffering from the round-off error because the integration points always take moderate values. In addition, because the unneeded eigenvalues $\lambda_n$ ($|\lambda_n| \ll 0.1$) that correspond to complex wave vectors $k_n$ with Im$(k_n) \gg 0$ are far away from the contour, the contamination of unneeded eigencomponents into the subspace will be prevented by the numerical integration.

\subsection{Sakurai-Sugiura method for nonlinear eigenvalue problem}
\label{sec:subsec3.2}

The original SS method \cite{Sakurai2003} is developed for finding the eigenvalues of the generalized eigenvalue problem that lies in the domain. With an extension for the nonlinear eigenvalue problem,\cite{Asakura2009} the SS method can also be applied to the EEP without loss of the desirable features of the original algorithm. Herein, we focus on the SS method for finding the eigenpairs of the EEP for $k$ inside the rectangular region, as shown in Fig.~\ref{fig:ring}(b).

The SS method consists of two steps. The first step is generating the subspace by contour integration. Let $\Gamma$ be a counter-clockwise contour along each side of the rectangular region that encloses the target eigenvalues $k_1,k_2,...,k_{\widehat m}$, where ${\widehat m}$ is the number of eigenvalues inside $\Gamma$. Then, the moment matrix $S_p (p=0,1,...,2N_{mm}-1)$ associated with the target eigenpairs is defined as
\begin{equation}
S_p = \frac{1}{2 \pi {\rm i} } \oint_{\Gamma} z^p [E-H(z)]^{-1}V {\rm d}z, 
\label{eq:moment}
\end{equation}
where $V$ is a ${N \times N_{rh}}$ nonzero arbitrary matrix. $N_{rh}$ and $N_{mm}$ are the number of right hand sides and the order of moment matrices, respectively. They are input parameters that are set so as to satisfy ${N_{rh}N_{mm}>{\widehat m}}$. We refer the reader to Ref.~\onlinecite{Futamura2013} for details on how to determine these parameters efficiently. Here,
\begin{equation}
H(k) = e^{-{\rm i}ka} H_{l,l-1}+H_{l,l}+e^{{\rm i}ka}H_{l,l+1}. 
\end{equation}
From a numerical viewpoint, $S_p$ is approximated by numerical integration as
\begin{equation}
S_p \approx \widehat S_p = \sum_{j=1}^{N_{int}}w_j z_j^p [E-H(z_j)]^{-1}V, 
\end{equation}
where $z_j$ and $w_j$ are quadrature points and weights, respectively, which are determined by a quadrature rule such as the Gauss-Legendre quadrature rule. To retain the numerical accuracy, we replace the momental weight $z_j^p$ with the shifted and scaled one $((z_j-\gamma)/\rho)^p$. With this modification, $\widehat S_p$ is rewritten as

\begin{equation}
\widehat S_p = \sum_{j=1}^{N_{int}}\frac{w_j}{\rho} \Big( \frac{z_j-\gamma}{\rho} \Big)^p [E-H(z_j)]^{-1}V.
\label{eq:Sp_hat}
\end{equation}
In this study, $\gamma$ and $\rho$ are set as $\gamma = 0.1/a$ and $\rho = \pi / a$, respectively. Note that the eigenvalues are also shifted and scaled; then, the original eigenvalues of EEP should be recovered as $k_n \leftarrow \gamma + \rho k_n$.

The second step is extracting the eigenpairs from $\widehat S_p$. Following the procedure given in Ref.~\onlinecite{Asakura2009}, we define the complex moment matrix $\widehat \mu_p = V^{\dagger}\widehat S_p$, and we extract the target eigenvalues by solving the $N_{mm} N_{rh}\ll N$ dimensional generalized eigenvalue problem
\begin{equation}
      \widehat T^< x = \tau \widehat Tx,
       \label{eq:gep_h}
\end{equation}
with $N_{mm} N_{rh} \times N_{mm} N_{rh}$ Hankel matrices $\widehat T$ and $\widehat T^{<}$ defined as
\begin{align}
        \widehat T^{<} = \left[
     \begin{array}{cccc}
              \widehat \mu_1 & \widehat \mu_2 & \cdots & \widehat \mu_{N_{mm}} \\
              \widehat \mu_2 & \widehat \mu_3 & \cdots & \widehat \mu_{N_{mm}+1} \\
                \vdots & \vdots & \ddots & \vdots \\
              \widehat \mu_{N_{mm}} & \widehat \mu_{N_{mm}+1} & \cdots & \widehat \mu_{2N_{mm}-1}
        \end{array}
        \right], 
\end{align}
and
\begin{align}
        \widehat T = \left[
        \begin{array}{cccc}
                \widehat \mu_0 & \widehat \mu_1 & \cdots & \widehat \mu_{N_{mm}-1} \\
                \widehat \mu_1 & \widehat \mu_2 & \cdots & \widehat \mu_{N_{mm}} \\
                \vdots & \vdots & \ddots & \vdots \\
                \widehat \mu_{N_{mm}-1} & \widehat \mu_{N_{mm}} & \cdots & \widehat \mu_{2N_{mm}-2}
        \end{array}
        \right].
\end{align}
From the viewpoint of numerical stability and efficiency, we compute the rank $\widehat{m}$ of $\widehat T$ by a singular value decomposition
\begin{equation}
        \widehat T 
        = [U_1, U_2] \left[
                \begin{array}{ll}
                        \Sigma_1 & O \\
                        O & \Sigma_2
                \end{array}
        \right] \left[
                \begin{array}{ll}
                        W_1^\dagger \\
                        W_2^\dagger
                \end{array}
        \right] 
        \approx U_1 \Sigma_1 W_1^\dagger.
        \label{eq:svd}
\end{equation}
Here the term $U_2 \Sigma_2 W_2$ is omitted because $\Sigma_2 \approx 0$. Upon substituting Eq.~\eqref{eq:svd} into Eq.~\eqref{eq:gep_h}, the EEP is reduced to an $\widehat{m}$-dimensional standard eigenvalue problem, that is,
\begin{equation}
        U_1^\dagger \widehat T^{<} W_1 \Sigma_1^{-1} y = \tau y,
        \label{eq:sep_h}
\end{equation}
where $y=\Sigma_1W_1^{\dagger}x$. The (approximated) eigenpairs are obtained as $(k_n,\phi_n) = (\gamma + \rho \tau, \widehat{S} W_1 \Sigma_1^{-1} y)$, where $\widehat{S} = [\widehat{S}_0,\widehat{S}_1, \dots, \widehat{S}_{N_{mm}-1}]$. In our algorithm, we use Eq.~\eqref{eq:sep_h} instead of Eq.~\eqref{eq:gep_h}. If there are too many eigenvalues inside the contour, the rectangular region should be divided into several subdomains to reduce the cost of solving Eq.~\eqref{eq:sep_h}. In this case, it might be better to set a subdomain to a quadrant in the complex $k$ plane because the number of eigenvalues located in each quadrant should be the same, owing to the quadruple relationship $(k_n,k_n^*,-k_n,-k_n^*)$.

\subsection{Reduction of computational cost of numerical integration}
\label{sec:subsec3.3}

In actual calculations, the numerical calculation of the contour integral in Eq.~\eqref{eq:moment} is an important issue because this procedure is a major part of the entire computation in the SS method. Eq.~\eqref{eq:moment} can be rewritten as the sum of four definite integrals:
\begin{equation}
S_p =S_{p}^{(1)}+S_{p}^{(2)}+S_{p}^{(3)}+S_{p}^{(4)},
\end{equation}
where 
\begin{eqnarray}
\label{eq:S_p1}
S_p^{(1)} &=& \frac{1}{ 2 \pi {\rm i} }  \int_{-\pi/a}^{\pi/a} \Big(x+ {\rm i} \frac{\ln \lambda_{min}}{a} \Big)^p  \Big[ E-H \Big( x+{\rm i}\frac{\ln \lambda_{min}}{a}\Big)\Big]^{-1} V {\rm d}x, \\
S_p^{(2)} &=& \frac{1}{2 \pi {\rm i} } \int_{\ln \lambda_{min}/a }^{- \ln \lambda_{min}/a } \Big(\frac{\pi}{a} + {\rm i}y \Big)^p \Big[E-H\Big(\frac{\pi}{a} + {\rm i}y \Big) \Big]^{-1} V { \rm i}{\rm d}y,\\
S_p^{(3)} &=& \frac{1}{ 2 \pi {\rm i}} \int_{\pi/a}^{-\pi/a} \Big(x -{\rm i} \frac{\ln \lambda_{min}}{a} \Big)^p \Big[E-H \Big(x-{\rm i}\frac{\ln \lambda_{min}}{a}\Big)\Big]^{-1}V {\rm d}x, \\
S_p^{(4)} &=& \frac{1}{ 2 \pi {\rm i} } \int_{-\ln \lambda_{min}/a }^{\ln \lambda_{min}/a } \Big(-\frac{\pi}{a}+{\rm i}y \Big)^p \Big[E-H\Big(-\frac{\pi}{a} + {\rm i}y \Big) \Big]^{-1}V { \rm i}{\rm d}y.
\label{eq:S_p4}
\end{eqnarray}
We use the $N_q$-point Gauss-Legendre quadrature rule to evaluate the definite integrals. Here, $N_q$ is a pair of polynomial orders, that is, $N_q=(N_{q1},N_{q2})$. Fig.~\ref{fig:ring}(b) shows $N_{q1}=6$ quadrature points along the Re($k$) axis and $N_{q2}=6$ quadrature points along the Im($k$) axis. The total number of quadrature points is $N_{int}=2N_{q1}+2N_{q2}$. At each quadrature point, we need to solve linear systems with multiple right-hand sides:
\begin{equation}
[E-H(z_j)]Y_j=V, \quad (j=1,2,...,N_{int}),
\label{eqn:sle2}
\end{equation}
where $z_j$ is the $j$-th quadrature point. By using the symmetry of the Hamiltonian matrix $H(k)$, the number of linear systems to be solved can be reduced to $N_{q1}+\frac{1}{2}N_{q2}$. If time-reversal symmetry holds, then $H_{l,l}=H_{l,l}^{\dagger}$ and $H_{l,l-1}=H_{l,l+1}^{\dagger}$, and we have 
\begin{eqnarray}
[E-H(z_j)]^{\dagger}&=& E-H(z_j^{\ast}). \label{eqn:symmetry1}
\end{eqnarray}
Eq.~\eqref{eqn:symmetry1} suggests that linear systems with Im$(z_j)>0$ are adjoints of linear systems with Im$(z_j)<0$. As noted in Appendix \ref{appendix:B}, the BiCG method can solve both systems simultaneously with very little additional computational cost. Furthermore, owing to the translational symmetry, $e^{{\rm i}ka}=e^{{\rm i}(ka+2\pi)}$ holds, and this leads to
\begin{eqnarray}
E-H\Big(-\frac{\pi}{a} + z_j \Big)  &=& E-H\Big(\frac{\pi}{a} + z_j \Big). \label{eqn:symmetry2}
\end{eqnarray}
It is clear from Eq.~\eqref{eqn:symmetry2} that the linear systems in $S_p^{(2)}$ and $S_p^{(4)}$ are the same; thus, we only need to solve either one of them. From the above, the numerical integration can be performed by solving the linear systems indicated by black dots in Fig.~\ref{fig:ring}(b).

Basically, first-principles electron transport calculations must be performed independently at each energy point. Therefore, the total computational cost for determining the generalized Bloch states is proportional to the number of energy points $N_{ene}$ if we solve the linear systems of Eq.~\eqref{eqn:sle2} independently. However, by using the shift-invariant property of the Krylov subspace, we can dramatically reduce the computational cost. The essence of this method is that we only solve the linear systems at the reference energy point and update the solutions at other energy points with moderate computational costs. This method is called the shifted BiCG method, and its details are presented in Appendix \ref{appendix:B}. Hereafter, we write a set of $N_{ene}$ energy points as simply $E=[E_1,E_2,...,E_{N_{ene}}]$. The algorithm for evaluating the self-energy matrices by using the SS method and shifted BiCG method is shown below.

\begin{flushleft}
{\bf Algorithm: SS method to evaluate $\Sigma_{L0}$ and $\Sigma_{R0}$}

{\it Input}$:N, M, N_{rh}, N_{m}, N_{int}, N_{ene}, \lambda_{min}, \rho,V \! \in \! \mathbb{C}^{N \times N_{rh}}$

{\it Output}$:\Sigma_{L0}$, $\Sigma_{R0} \in \mathbb{C}^{M \times M}$

{\bf 1.} Set a rectangular contour in the $k$ plane for a given $\lambda_{min}$.

{\bf 2.} Set $(w_j,z_j)_{1 \le j \le N_{int}}$ by the $N_q$-point Gauss-Legendre quadrature rule. 

{\bf 3.} Set energy points $E=[E_1,E_2,...,E_{N_{ene}}]$.

{\bf 4.} Compute $[E-H(z_j)]Y_j=V$ by the shifted BiCG method for $j=1,2,...,N_{int}$.
\end{flushleft}

\noindent
{\bf Note.} Later processes are performed independently at each energy point.

\begin{flushleft}
{\bf 5.} Compute $\widehat S_p$ by Eq.~\eqref{eq:Sp_hat} and set $\widehat \mu_p=V^{\dagger} \widehat S_p$ for $p= 0,1,...,2N_{mm}-1$. 

{\bf 6.} Set $[\widehat T]_{ij} \! = \! \widehat \mu_{i+j-1}$ and $[\widehat T^<]_{ij} \!=\! \widehat \mu_{i+j-2}, 1 \! \le \! i,j \! \le \! N_{mm}.$

{\bf 7.} Approximate $\widehat T \approx U_1 \Sigma_1 W_1^\dagger$ by singular value decomposition.

{\bf 8.} Solve $U_1^\dagger \widehat{T}^{<} W_1 \Sigma_1^{-1} y = \tau y$.

{\bf 9.} Extract eigenpairs $(k_n,\phi_n) = (\gamma + \rho \tau, \widehat{S} W_1 \Sigma_1^{-1} y)$.

{\bf10.} Construct ${Q}_{Ll}^{-},{Q}_{Rl}^{+},{\widetilde Q}_{Ll}^{-},{\widetilde Q}_{Rl}^{+}$ using Eqs.~\eqref{eq:Bloch_mat1}-\eqref{eq:Bloch_mat4} for $l=0,1$.

{\bf11.} Construct ${\Sigma}_{L0}$ and ${\Sigma}_{R0}$ using Eq.~\eqref{eq:SE}.
\end{flushleft}

\section{Numerical tests}
\label{sec:sec4}
In this section, we demonstrate the numerical accuracy, robustness, and efficiency of our method through a series of test calculations. KS Hamiltonian matrices are obtained from the real-space pseudopotential DFT code RSPACE.\cite{Ono1999,ICP} All calculations in this and the later sections are performed by using the local density approximation \cite{Perdew1981} and norm-conserving pseudopotentials proposed by Troullier and Martins.\cite{Troullier1991} In all cases, $\Gamma$-point sampling in the two-dimensional Brillouin zone is used. Unless noted otherwise, we used a fourth-order finite-difference approximation for the Laplacian operator and a grid spacing of 0.2 \AA.

\subsection{Accuracy of eigenpairs inside $\Gamma$}
\label{sec:subsec4.1}

\begin{figure*}
\begin{center}
\includegraphics[width=160mm]{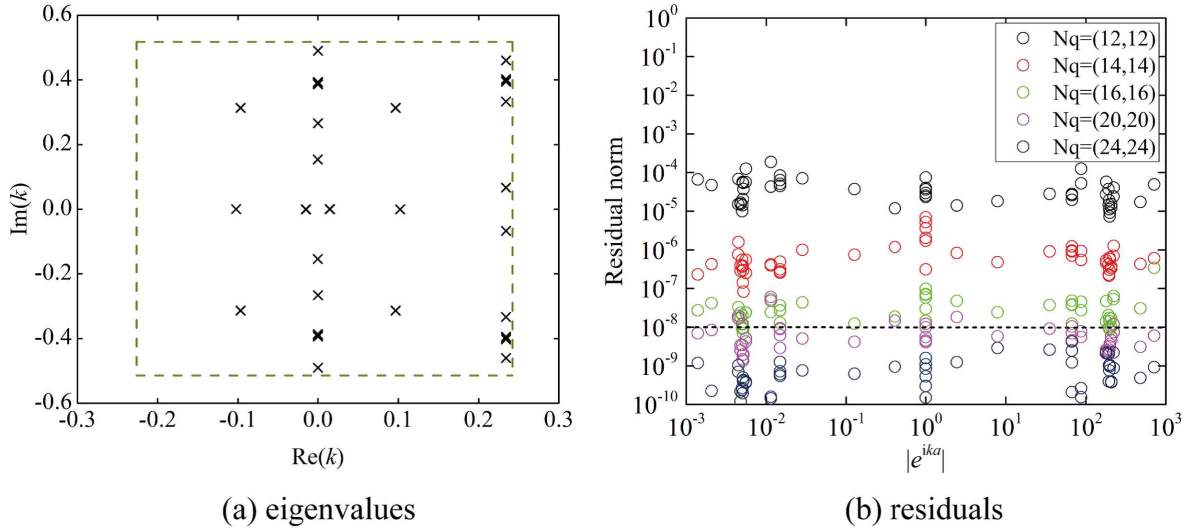}
\caption{Numerical results for fcc Au bulk at the Fermi energy of (a) distribution of eigenvalues within the domain enclosed by $\Gamma$ and (b) residuals $||[E-H(k_n)]\phi_n||_{2}$ when varying the order of the Gauss-Legendre quadrature rule, $N_q=(N_{q1},N_{q2})$. The number of target eigenvalues that do not include spurious eigenpairs is 54. The plots clearly show that the positions of the eigenpairs are almost unchanged and that the accuracy is straightforwardly improved by increasing $N_q$. Convergence is achieved at $N_q=(24,24)$: all target eigenvalues are below the convergence criterion, as indicated by the broken line.}
\label{fig:conv_integral}
\end{center}
\end{figure*}

First, we confirm the accuracy of the \textbf{Algorithm} described in Sec. \ref{sec:subsec3.3}. In addition to the Hamiltonian matrix and inner radius $\lambda_{min}$, the SS method requires several other parameters: order of Gauss-Legendre quadrature rule $N_q=(N_{q1},N_{q2})$, number of right-hand sides $N_{rh}$, and order of moment matrices $N_{mm}$. It is essential to select these parameters appropriately to make the algorithm robust and efficient. Among these parameters, $N_{rh}$ and $N_{mm}$ are also used in the conventional SS method, and their effects on numerical errors have been studied elsewhere. Thus, we expect that general principles can be applied to $N_{rh}$ and $N_{mm}$.\cite{z-Pares} Care must be taken when selecting $N_q$ because the SS method features an ellipsoid-type contour, with numerical integration performed using the trapezoidal rule. Because the numerical integration method for the SS method using a rectangular-type contour has not been proposed, the Gauss-Legendre quadrature rule is employed in this study. Thus, we demonstrate the selection of $N_q$ by monitoring the residuals defined as $||[E-H(k_n)]\phi_n||_{2}$. The generalized Bloch states are normalized such that $||\phi_n||_2=1$. Note that we remove eigenpairs whose residuals are greater than $10^{-1}$ or located outside $\Gamma$ as spurious eigenpairs. Here, we consider the fcc Au bulk with 18 atoms whose transport direction is parallel to the $\langle 111 \rangle$ direction. We set $N_{mm}=8$, $N_{rh}=16$, and $\lambda_{min}=0.001$. The criterion of the singular value decomposition and the shifted BiCG method is set to $10^{-15}$. Fig.~\ref{fig:conv_integral}(a) shows the distribution of the eigenvalues when $N_{q}=(24,24)$. It should be noted that the obtained eigenvalues are pairwise, that is, $(k_i,k_j) \approx (k_i,k_i^{\ast})$, and the number of eigenvalues is unchanged irrespective of the selection of $N_{q}$. In Fig.~\ref{fig:conv_integral}(b), the residuals $||[E-H(k_n)]\phi_n||_{2}$ are plotted as a function of $N_{q}$. The accuracy of the obtained eigenpairs is uniformly improved by increasing $N_q$, and convergence can be achieved with a relatively small number of quadrature points (convergence criterion is set to $10^{-8}$). It should be noted that better accuracy can also be obtained by increasing $N_{rh}$ without using large values of $N_q$. Further details on how to improve the accuracy of the nonlinear SS method are presented in Ref.~\onlinecite{Yokota2013}.

\subsection{Robustness of the algorithm}
\label{sec:subsec4.2}

\begin{figure*}
\begin{center}
\includegraphics[width=160mm]{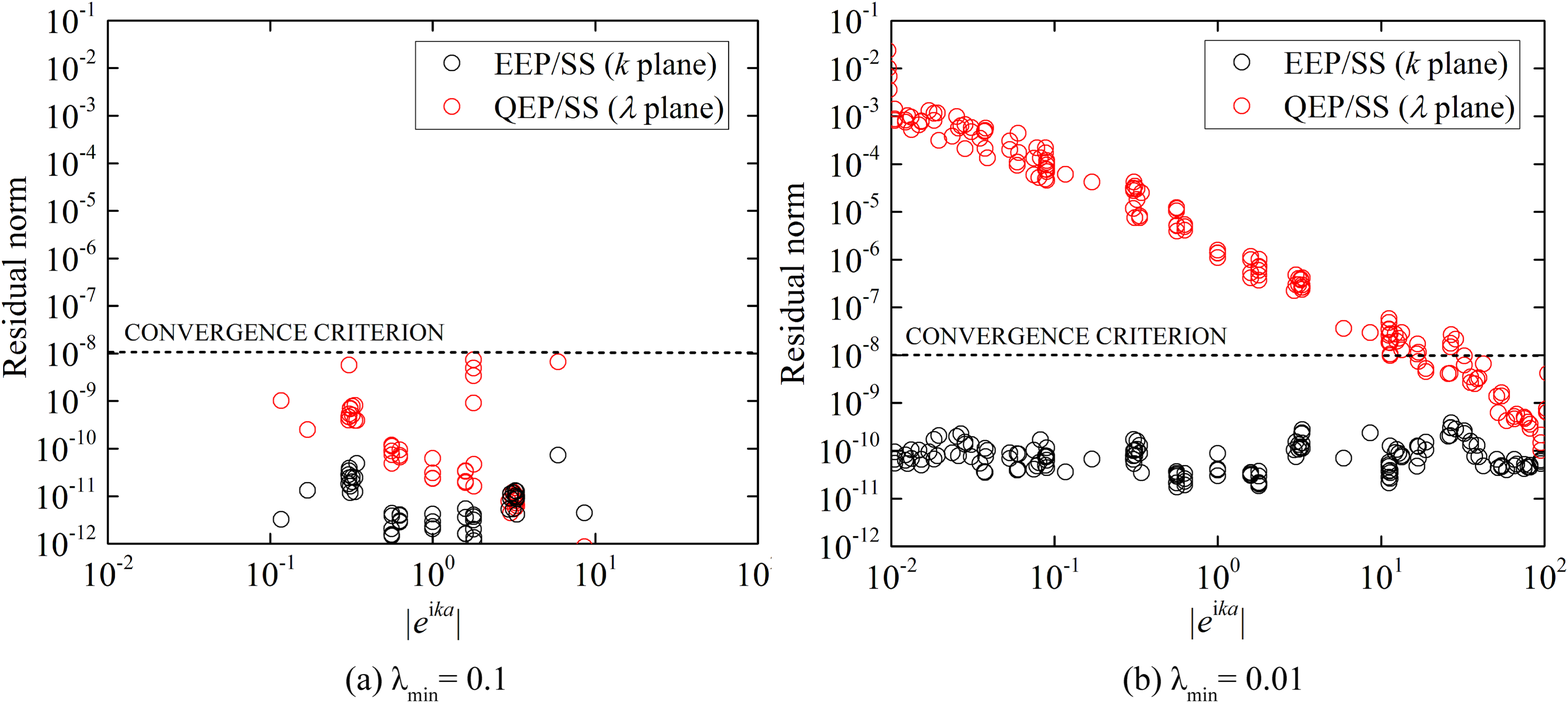}
\caption{Residuals $||[E-H(k_n)]\phi_n||_{2}$ for (6,6)CNT at the Fermi energy calculated on the $k$ plane (EEP/SS) and $\lambda$ plane (QEP/SS). (a) For $\lambda_{min}=0.1$, the number of eigenvalues that do not include spurious eigenpairs is 52. (b) For $\lambda_{min}=0.01$, the number of eigenvalues that do not include spurious eigenpairs is 154. In both cases, the EEP/SS method uses $N_q=(24,24)$ as the order of the Gauss-Legendre quadrature rule; by contrast, the QEP/SS method uses the trapezoidal rule with the number of quadrature points being 256 per circle in Fig.~\ref{fig:ring}(a). The other parameters are kept the same.}
\label{fig:QEP-NEP}
\end{center}
\end{figure*}

In Sec.~\ref{sec:subsec3.1}, we stated that our method is more robust than previous methods that solve Eq.~\eqref{eq:QEP} as a QEP for $\lambda(=e^{{\rm i}k_na})$ when $\lambda_{min}\ll 0.1$. To demonstrate the robustness of our method, we first compare the eigenvalues and residuals computed on the $k$ plane in Fig.~\ref{fig:ring}(b) with those computed on the $\lambda$ plane in Fig.~\ref{fig:ring}(a). For comparison, we also apply the algorithm proposed in Ref.~\onlinecite{Iwase2017} for solving the QEP for $\lambda$ by using the contour along the ring-shaped region in Fig.~\ref{fig:ring}(a). Here, we use the (6,6)CNT with 24 atoms whose transport direction is parallel to the channel direction. The input parameters of the SS method are set to $N_{mm}=4,N_{rh}=128$, and the criterion of the singular value decomposition and the shifted BiCG method is set to $10^{-15}$. The order of the Gauss-Legendre quadrature rule is $N_q=(24,24)$ in the $k$ plane computation. Instead, we use the trapezoidal rule to approximate the contour integrals in Fig.~\ref{fig:ring}(a), with the number of quadrature points being 256. Fig.~\ref{fig:QEP-NEP} shows the eigenvalues and residuals calculated on the $k$ plane and $\lambda$ plane for (a) $\lambda_{min}=0.1$ and (b) $\lambda_{min}=0.01$. In both cases, all residuals computed on the $k$ plane are far below the convergence criteria irrespective of the eigenvalues. It is important to note that a satisfactory result is also obtained when setting $\lambda_{min}$ = $0.001$. On the contrary, when $\lambda_{min}=0.01$, the accuracy of the solutions computed on the $\lambda$ plane is quite poor, and the accuracy is seen to degrade from the outer circle toward the inner circle.

As mentioned in Sec.~\ref{sec:subsec3.1}, this round-off error of the $\lambda$-plane computation can be reduced by adding some circles between the outer and the inner circles. However, an additional computational cost is incurred when we use the BiCG method for solving Eq.~\eqref{eqn:sle2}. If we use the BiCG method as the solver, it is difficult to set a reasonable number of right-hand side $N_{rh}$ owing to a variation in the eigenvalue distribution inside the ring-shaped region. For example, when $\lambda_{min}=0.001$, roughly 10 circles should be added between the inner and the outer circles to make the round-off error sufficiently small. In the case of the fcc Au bulk shown in Fig.~\ref{fig:conv_integral}(a), a maximum of 33\% of eigenvalues are in a sliced ring. Furthermore, the risk of the computation failing also increases because the distribution of the eigenvalues located inside the sliced rings is unknown in advance and is model-dependent. 

\subsection{Serial performance}
\label{sec:subsec4.3}

In this subsection, we experimentally evaluate the serial performance of our method for the eigenvalue problem arising from the computation of self-energy matrices. To demonstrate  speed-ups, we compare the computational time of our method with that of the OBM method, which is categorized as a WFM method. Although continuous improvements\cite{Tsukamoto2014,Ono2012,Iwase2015,Ono2016} have been made after the first study of the OBM method, the computation of the first and last $r$ columns of $(E-H_{l,l})^{-1}$ and the $2r$-dimensional generalized eigenvalue problem is still required. In this study, the matrix inversion is calculated using the CG method,\cite{CG} and the generalized eigenvalue problem is solved by the optimized LAPACK routine ZGGEV and SS method.\cite{Tsukamoto2014} It should be noted that another method based on a real-space grid approach proposed by Khomyakov $et~al$.\cite{Khomyakov2004} is not considered here because its computational procedure and cost are almost the same as those of the OBM method. In addition, popular methods\cite{Sancho1984,Sancho1985,Sanvito1999,Rungger2008,Sorensen2008} used in the NEGF method are also excluded from consideration because they involve the inversion of very dense matrices with the size of the real-space grids in the unit cell of the electrode.

Table~\ref{table:runtime} shows the breakdown of the profiling results in various test systems. All calculations are performed on a two-socket Intel Xeon E5-2667v2 with 16 cores (3.3 GHz) and 256 GB of system memory. 4 MPI processes and 4 OpenMP threads are assigned to the CPU. The parameter $\lambda_{min}$ is set to $0.1$, as in Ref.~\onlinecite{Sorensen2008}. The input parameters of the proposed method are set as $N_{mm}=8$, $N_q=(24,24)$, and $N_{ene}=100$, and the criterion of the singular value decomposition and shifted BiCG method is set to $10^{-15}$. The number of right-hand sides $N_{rh}$ is set such that all residuals are less than $10^{-8}$. Equidistant energy points are chosen in the interval $E-E_F \in [-1,1]$ eV, where $E_F$ is the Fermi energy. The CPU times of the proposed method listed in the sixth column in Table~\ref{table:runtime} represent the average calculation times for 100 energy points. For the SS method used in the OBM method, we employ the trapezoidal rule with the number of quadrature points being 32 per circle in Fig.~\ref{fig:ring}, which is the default value used in z-Pares, \cite{z-Pares} and other parameters are set as the same in the proposed method. The CPU times including both CG and ZGGEV (SS) contributions listed in the fourth (fifth) column are evaluated from the computation at the Fermi energy owing to the limitation of computational resources. The CPU times of these two reference methods can be reduced by employing the shifted CG method instead of the standard CG method and using smaller number of quadrature points. To use the computer resources efficiently, we need to optimize the parameters for the shifted CG and SS methods, which depend on the test systems. Since the usage of uneven parameters become an obstacle to demonstrate the characteristic advantage of the proposed method, the standard CG method is employed and the number of quadrature points is set to be the default value in z-Pares. It was verified that the proposed method is faster than the reference methods in order by one-, two-, and three-dimensional systems. The computational cost of the proposed method scales as $O(NN_{itr}N_{int}N_{rh})$, while that of the CG/SS method does as $O(r^3N_{int})$, where $N_{itr}$ is the number of iterations for the shifted BiCG method and increases linearly or more with the common logarithm of $1/\lambda_{min}$. The number of target eigenvalues generally decreases against its matrix size as the dimension of the systems becomes smaller, indicating that the proposed method is much more efficient in the low dimensional systems because $N_{rh}$ can be set to be small number.

Another important way to reduce the computational cost is to use different numbers of processors, that is, to use parallel computing. The scalability of the SS method based on a real-space grid approach has already been investigated in our preliminary work.\cite{Iwase2017} Here, we only mention that the parallel performance of the proposed method is superior to that of conventional eigenvalue solvers because of the hierarchical parallelism of the contour integral approach and the domain-decomposition technique for the sparse Hamiltonian matrix $H(k)$.

\begin{table*}
\begin{threeparttable}
\caption{CPU times in hours for computing the eigenvalue problems arising from the self-energy computations for various electrode materials. Here, $N$ is the size of the Hamiltonian matrix, and $2r$ is the number of nontrivial solutions of Eq.~\eqref{eq:QEP}. $N_{rh}$ is the number of right-hand sides used in the SS method. The CPU times of the proposed method (this work) are averaged by the computation times at 100 different energy points between $E_F-1$~eV and $E_F+1$~eV, where $E_F$ is the Fermi energy. On the other hand, the CPU times of the OBM method (CG/ZGGEV,CG/SS) are measured only at the Fermi energy owing to the limitation of computational resources.}
{
\label{table:runtime}
\begin{tabularx}{170mm}{CCCCCCC}
\hline \hline
Material & $N$ & $2r$ & $N_{rh}$ & CG/ZGGEV & CG/SS & This work \\
\hline
Au chain \tnote{a} & 64896 & 21632 & 8 & 13.43 & 4.11 & 0.01 \\
Al(100) wire \tnote{b} & 153600 & 51200 & 16 & 190.06 & 28.96 & 0.08 \\
(6,6)CNT \tnote{c}& 62208 &  41472 & 32 & 115.52 & 13.15 & 0.12 \\
(10,10)CNT \tnote{c}& 172800 &  115200 & 64 & -\tnote{d} & -\tnote{d} & 1.20 \\
Graphene \tnote{e}& 14336 & 7168 & 8 & 0.60 &  0.13 & 0.00 \\
Silicene \tnote{f}& 110592 & 24576 & 16 &  26.63 & 5.66 & 0.09 \\
Au(111) bulk \tnote{g}& 34560 & 7680 & 8 & 1.14  & 0.66 & 0.01\\
\hline \hline
\end{tabularx}
}
\begin{tablenotes}
\item[a] Geometry description is found in the transmission calculation of the Au atomic chain with a CO molecule. See Sec.~\ref{sec:sec5}. 
\item[b] Geometry description is found in Ref.~\onlinecite{Brandbyge2002}.
\item[c] Ideal armchair (n,n) carbon nanotube with C-C bond length of 1.42 \AA.
\item[d] Calculation fails owing to a memory exhaustion error.
\item[e] Graphene with four atoms whose transport direction is along the armchair direction and C-C bond length is 1.42 \AA. 
\item[f] Geometry description is found in the transmission calculation of the free-standing silicene. See Sec.~\ref{sec:Application}.
\item[g] Geometry description is found in Ref.~\onlinecite{Strange2008}.
\end{tablenotes}
\end{threeparttable}
\end{table*}

\section{Transmission calculation}
\label{sec:sec5}

In this section, we present the transmission calculations for Au atomic chain with a CO molecule. We chose this system because transport properties have been investigated extensively using other methods.\cite{Calzolari2004,Strange2008} To validate the accuracy of our method for electron transport calculations, we study the effect of excluding rapidly decaying evanescent waves on the zero-bias transmission calculation, and we compare this result with those obtained using other methods. In all calculations, the KS Hamiltonian matrices in the transition region are obtained from the DFT calculation under periodic boundary conditions, and the scattering states in real-space grids are calculated using the IOBM method.\cite{Kong2006,Kong2007} 

\begin{figure*}
\begin{center}
\includegraphics[width=100mm]{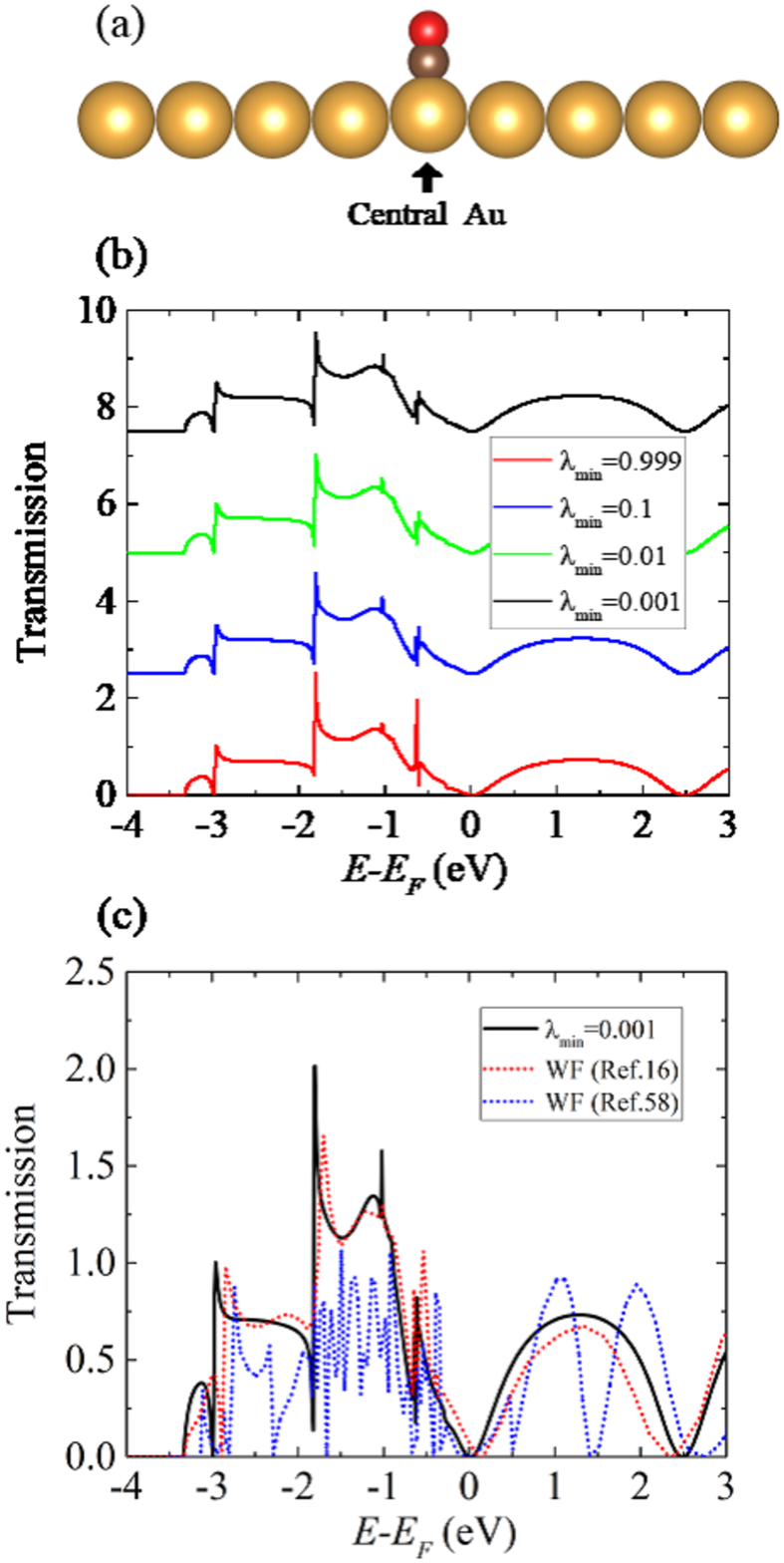}
\caption{(a) Transition region of Au atomic chain with CO adsorption. Au, C, and O atoms are represented as gold, brown, and red balls, respectively. (b) Transmission spectra obtained using self-energy matrices calculated by the proposed method with four different $\lambda_{min}$ values: 0.999 (red line), 0.1 (blue line), 0.01 (green line), and 0.001 (black line). For clarity, transmission spectra are shifted by the amount of 2.5 with respect to the original values in descending order of the legend. (c) The transmission spectrum obtained with the proposed method (black line). The results obtained with the maximally localized Wannier function (WF) of Ref.~\onlinecite{Strange2008} (red dashed line) and Ref.~\onlinecite{Calzolari2004} (blue dashed line) are also shown in (c).}
\label{fig:Au-CO}
\end{center}
\end{figure*}

We present the transmission calculation of the Au atomic chain with a CO molecule. Prior to our work, transmission calculations for this system have been done by Calzolari $et~al.$\cite{Calzolari2004} and Strange $et~al..$\cite{Strange2008} Interestingly, both groups reported relatively different transmission curves, even though they employed the same methodology combined with the Green's function method and maximally localized Wannier function. As stated in Ref. \onlinecite{Strange2008}, the disagreement might be related to the manner of construction of the tight-binding Hamiltonian, but it is still unresolved as to which one is correct. Thus, we revalidate the transmission calculation of the Au atomic chain with a CO molecule using the real-space grid method. Fig.~\ref{fig:Au-CO}(a) shows the atomic structure of the Au atomic chain with a CO molecule. The transition region is a rectangular box of $12.00 \times 12.00 \times 26.10$ \AA$^3$, and electron transport occurs along the $z$ direction. As in Refs.~\onlinecite{Calzolari2004,Strange2008}, the bond lengths are set as $d_{{\rm Au-Au}}=2.90$ \AA, $d_{{\rm Au-C}}=1.96$ \AA, and $d_{{\rm C-O}}=1.15$ \AA, and the Au atom attached to CO is shifted toward CO by $0.20$ \AA. The left and right electrodes are infinite Au chains with two atoms in the unit cell. A grid spacing of 0.23 \AA~is used in the real-space grid calculation. Fig.~\ref{fig:Au-CO}(b) shows the transmission spectra obtained using the self-energy matrices calculated by the proposed method with $\lambda_{min}=0.999,0.1,0.01,0.001$. In all calculations, we reproduce the main features in Fig.~2 of Ref.~\onlinecite{Strange2008}: (i) the drop in the transmission at the Fermi energy that originates from resonant scattering by CO adsorption, (ii) the single broad peak at $E - E_F \in [0,2]$ eV, and (iii) spiky peaks at $E - E_F \in [-4,0]$ eV. As the $\lambda_{min}$ value decreases, the transmission spectra rapidly converge toward the correct values, and visible differences are not observed when $\lambda_{min} \le 0.01$. In Fig.~\ref{fig:Au-CO}(c), the real-space grid calculation shows the qualitatively good agreement with the curve of Ref.~\onlinecite{Strange2008}; however, we found some discrepancies between the real-space grid calculation and the curve of Ref.~\onlinecite{Calzolari2004}. 

\begin{figure*}
\begin{center}
\includegraphics[width=120mm]{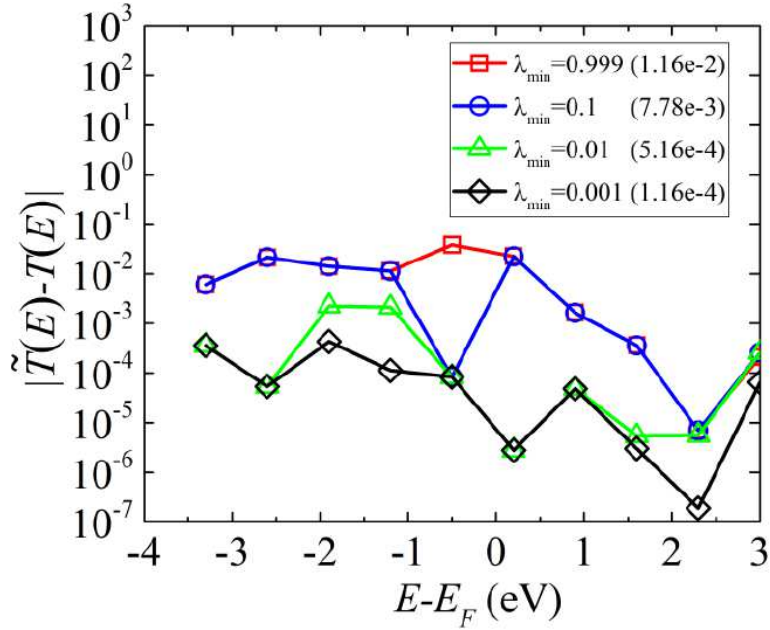}
\caption{Differences between transmission spectra using approximated and exact self-energy matrices. The overall deviation is indicated in the parentheses following the legends.}
\label{fig:Au-CO_error}
\end{center}
\end{figure*}

To investigate the accuracy of the approximated self-energy matrices calculated using the proposed method further, we measure the deviation from the exact transmission probability,
\begin{equation}
\Delta = \frac{1}{E_1+E_2}\int_{E_F-E_2}^{E_F+E_1} |\widetilde T(E)-T(E)|dE,
\end{equation}
where $\widetilde T(E)$ and $T(E)$ are the transmission probabilities obtained using the approximated and exact self-energy matrix, respectively. The exact self-energy matrices are calculated using the continued-fraction method.\cite{Ono2016} Fig.~\ref{fig:Au-CO_error} shows plots of the differences $|\widetilde T(E)-T(E)|$ and deviations $\Delta$. Here, the results at $E=E_F-4$~eV are omitted because the propagating wave does not exist. We find that $|\widetilde T(E)-T(E)|$ shows step-like reductions when using smaller values of $\lambda_{min}$. This is because the number of generalized Bloch states changes discretely with decreasing $\lambda_{min}$. The deviations $\Delta$ are as follows: $1.16 \times 10^{-2}$ ($\lambda_{min}=0.999$), $7.78 \times 10^{-3}$  ($\lambda_{min}=0.1$), $5.16 \times 10^{-4}$ ($\lambda_{min}=0.01$), and $1.16 \times 10^{-4}$ ($\lambda_{min}=0.001$). Note that our results are an order of magnitude different from the values reported in Ref.~\onlinecite{Sorensen2008} for the same value of $\lambda_{min}$. This is because that the self-energy matrices used in our transport calculation are not same as used in Ref.~\onlinecite{Sorensen2008}. We define the self-energy matrices at $L0$ and $R0$ regions in Fig.1 using the generalized Bloch states and use them in the transport calculations, while the previous study first define the self-energy matrices using the generalized Bloch states at $L1$ and $R1$ regions, and calculate the self-energy matrices at $L0$ and $R0$ regions by the recursion method. Although both approaches are equivalent when the self-energy matrices are exact, the later is more accurate than the former when the cut off for evanescent waves is introduced because the later uses the self-energy matrices defined at one layer deep inside the electrodes.

\section{Application}
\label{sec:Application}
Silicene, which is a two-dimensional honeycomb structure of Si atoms, is a promising candidate for future nanoelectronic devices due to its unique electronic structures, as represented by a zero-gap semiconductor with Dirac cone.\cite{Takeda1994,Lebegue2009} In fact, a silicene field effect transistor (FET) using the transfer-fabrication process was recently reported.\cite{Tao2015} However, the measured mobility values of silicene FET are considerably lower than the theoretical calculation\cite{Li2013} by an order of magnitude, and grain boundary scattering has been proposed as a possible cause. Despite the demand for the detailed information on the electron scattering at the grain boundaries of silicene, the electron transport behavior across the grain boundary of silicene is not well understood. This is because the fabrication of silicene is still challenging, especially on a dielectric substrate. 

\begin{figure*}
\begin{center}
\includegraphics[width=160mm]{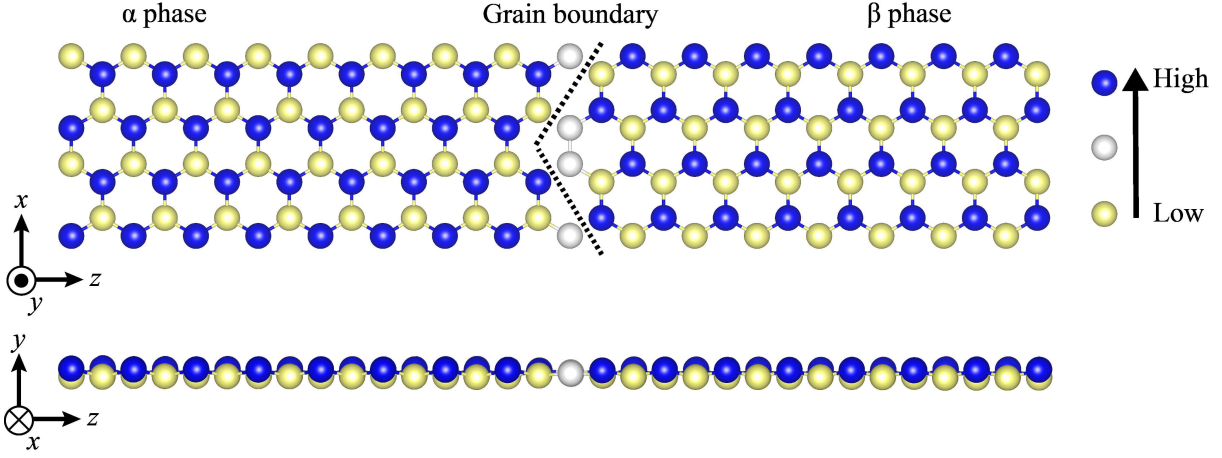}
\caption{Optimized interface structure of silicene with $\alpha$-$\beta$ interface.}
\label{fig:silicene_model}
\end{center}
\end{figure*}

Unlike graphene, it is well known that silicene forms a low-buckled structure, which leads to two energetically equivalent geometrical phases, whose buckling directions are opposite to each other, as shown in Fig.~\ref{fig:silicene_model}. Following the notations in Ref.~\onlinecite{Lima2013}, we call these phases the $\alpha$ and $\beta$ phases, respectively. In this study, we present the first-principles analysis of the transport properties of the free-standing silicene sheet across the interface between the $\alpha$ and $\beta$ phases. To keep the focus on application of our methodology to the transport calculations rather than on comprehensive understanding of the scattering mechanism in silicene, attention is paid only to the grain boundary between the $\alpha$ and $\beta$ phases along the armchair direction.

Initially, we perform the relaxation of the interface structure using a grid spacing of $0.21$ \AA~ and a $4\times1\times1$ $k$-point sampling on Brillouin zone. The interface model is constructed with a 256-atom supercell using a value of 2.27 \AA~ for the Si-Si bonding length. To avoid the spurious interaction between silicene layers, a vacuum region of 10 \AA~is introduced in the simulation cell. The interface structure is relaxed until the residual forces become lower than 0.003~eV/\AA. The relaxed geometrical structure is shown in Fig.~\ref{fig:silicene_model}. The reconstruction of chemical bonds at the interface does not occur, but instead the rearrangement of the out-of-plane dislocation is observed at the interface. This result is in agreement with previous theoretical work.\cite{Lima2013} We subsequently perform the transport calculations along the $z$ direction using the self-energy matrices evaluated with $\lambda_{min}=0.01$. The transition region contains 128 atoms and the $\Gamma$ point in the transverse direction is used in the transport calculation.

\begin{figure*}
\begin{center}
\includegraphics[width=120mm]{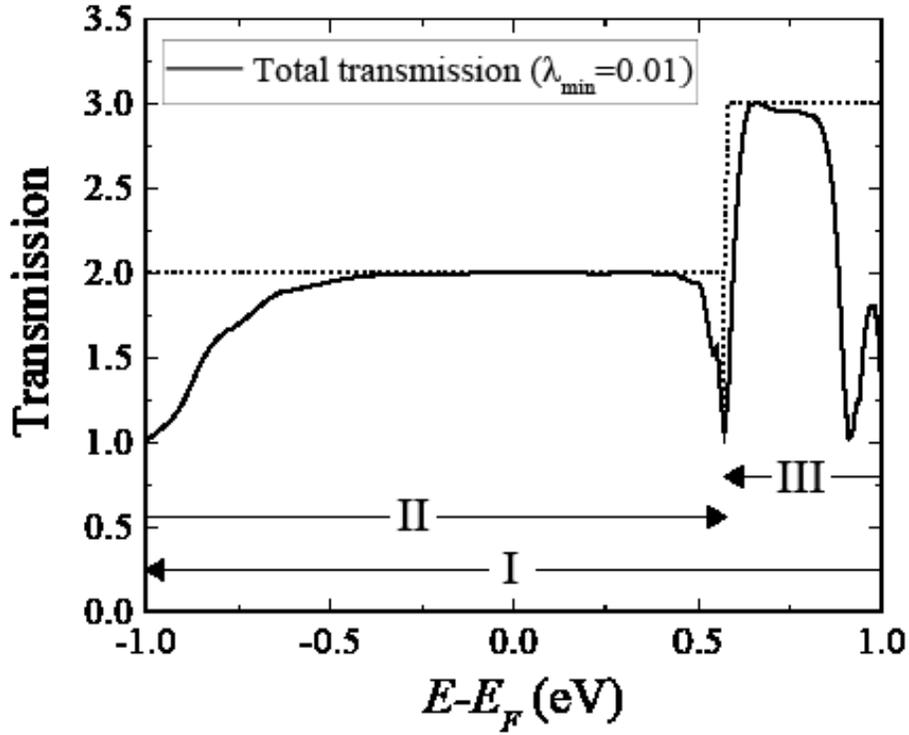}
\caption{Effects of the $\alpha$-$\beta$ interface on the transmission spectra. The solid line is the result of real-space grid calculation using the self-energy matrices obtained with $\lambda_{min}=0.01$. Empty dots denote the transmission spectrum without the defect.}
\label{fig:silicene_trans}
\end{center}
\end{figure*}

\begin{figure*}
\begin{center}
\includegraphics[width=120mm]{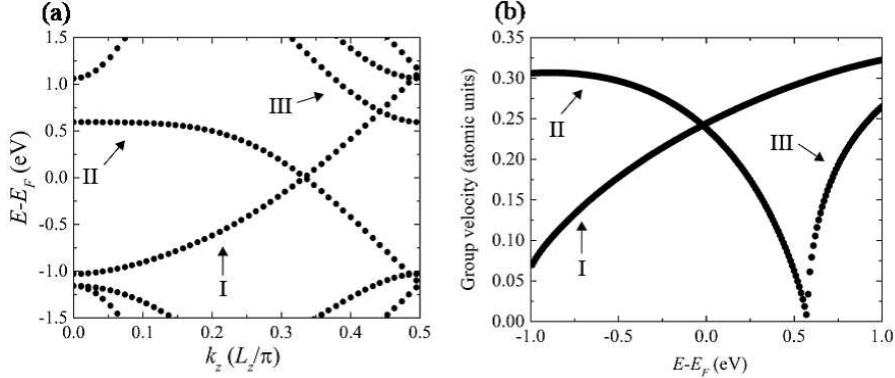}
\caption{(a) Band structure and (b) group velocity of silicene.}
\label{fig:silicene_velo_band}
\end{center}
\end{figure*}

The total transmission is shown in Fig.~\ref{fig:silicene_trans}. A feature of immediate interest is that the transmission at the Fermi energy is unchanged for a pristine silicene; that is, this type of grain boundary does not scatter incoming electrons at this energy. On the other hand, we found three transmission dips below and above the Fermi energy in Fig.~\ref{fig:silicene_trans}. Aiming to understand the origin of such dips, we also plot the group velocity of the incident electrons and the band structure of silicene in Fig.~\ref{fig:silicene_velo_band}. The important bands near the Fermi energy are labeled by I, II and III, and they contribute to the transmission in $[-1.0,1.0]$~eV, $[-1.0, 0.57]$~eV, and $[0.57, 1.0]$~eV, respectively. Note that the III band is doubly degenerated. At $E=E_F-1.0$ and $0.57$~eV where the transmission dips are observed, the group velocities go to zero with opening or closing the channel in bands. The scattering where the group velocity becomes zero is understood by the one-dimensional tight-binding model with a single impurity. According to Eq.~(75) in Ref.~\onlinecite{Khomyakov2005}, it is easy to show that the transmission probability becomes zero when the group velocity becomes zero, which indicates that the perturbation of the potential induced by the geometrical disorder at the silicene interface causes the scattering near the band edge. In addition, we numerically examine the scattering at the band edge using the one-dimensional Kronig-Penny model and observe the strong scatterings at the band edge (see, Appendix B.)

\begin{figure*}
\begin{center}
\includegraphics[width=80mm]{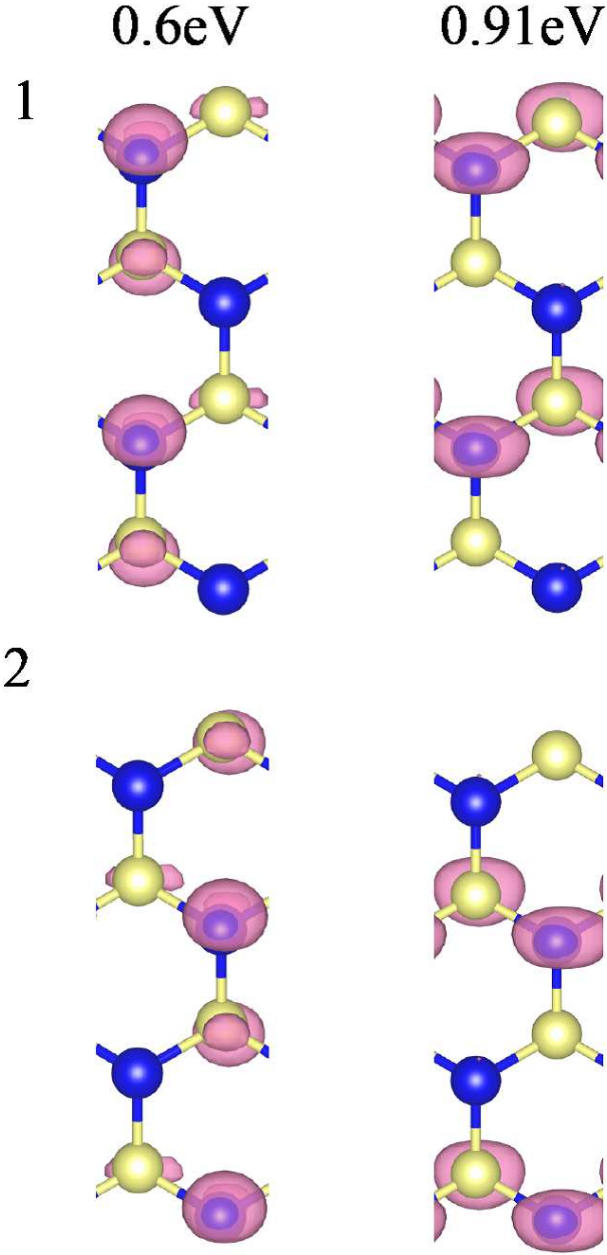}
\caption{Charge densities of two bulk modes in III band for left electrode at $E=E_F+0.6$ and $0.91$ eV.}
\label{fig:silicene_bulkmode}
\end{center}
\end{figure*}

\begin{figure*}
\begin{center}
\includegraphics[width=120mm]{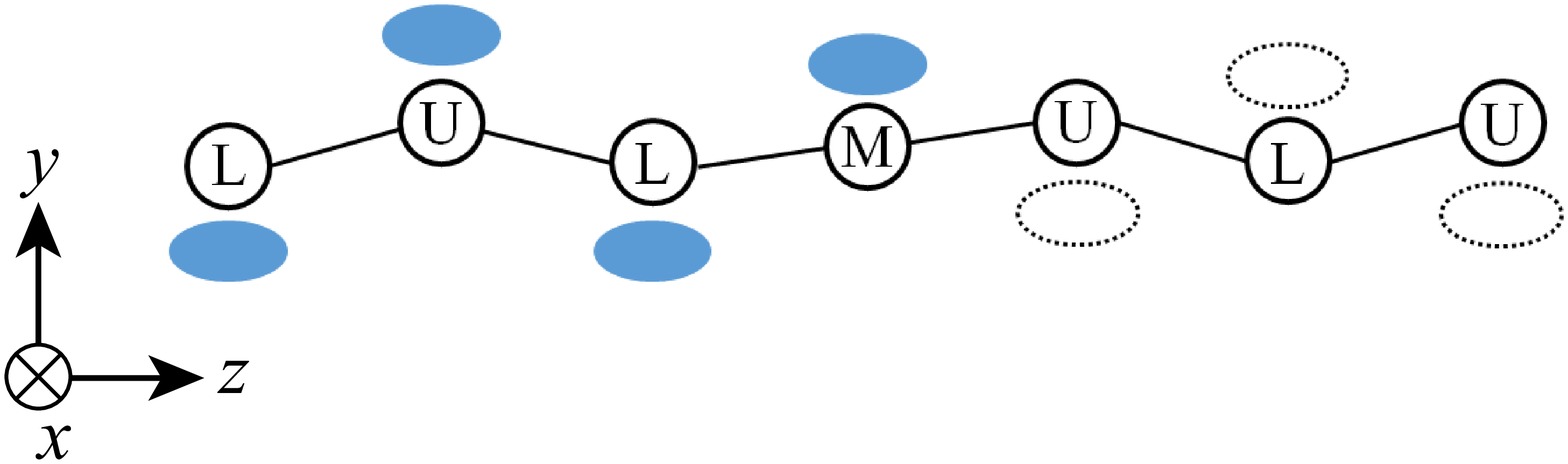}
\caption{An illustration of the scattering of silicene at $E=E_F+0.91$ eV. The symbols L, M, and U represent the lower buckled, non-bucked, and upper buckled atoms, respectively.}
\label{fig:silicene_scat}
\end{center}
\end{figure*}

We next consider the origin of the dip at $E=E_F+0.91$ eV, where two bulk modes in III band are completely reflected at the interface. To obtain the more detailed information about the scattering, we plot the charge densities of two bulk modes for left electrode at $E=E_F+0.6$ and $0.91$ eV. As seen from Fig.~\ref{fig:silicene_bulkmode}, the charge densities of two channels at $E=E_F+0.6$~eV distribute on inner and outer sides of silicene atoms, while they turn to concentrate on the only outer side at $E=E_F+0.91$~eV. By expanding the result for left electrode to the interface, the new insight of the scattering is obtained. Fig.~\ref{fig:silicene_scat} illustrates the scattering of the incident electron coming from the left electrode at $E=E_F+0.91$~eV. The bulk modes distribute around the outer side of the silicene atoms in both right and left electrodes, however, the scattering states in the right electrodes will be inner side of the silicene atoms because the buckling of the silicene is reversed. Therefore, the scattering states which come from the left electrode hardly connect with the bulk modes in right electrodes, which leads the transmission reduction at $E=E_F+0.91$~eV.

\section{Conclusion}
\label{sec:sec6}
This study proposes a robust and efficient method for evaluating the self-energy matrices of electrodes and provides its implementation for a real-space grid approach using a higher-order finite-difference scheme with nonlocal pseudopotentials. Considering that most generalized Bloch states decay rapidly, the transmission probabilities of nanoscale systems can be determined by using a comparatively smaller number of propagating and moderately decaying waves with practically sufficient accuracy. To obtain such physically important waves efficiently, a contour integral eigensolver based on the SS method combined with the shifted BiCG method is developed. Because the developed method does not involve any explicit computations requiring the inversion of the Hamiltonian matrix, the computational time and memory requirement are reduced.

Sec. \ref{sec:sec4} demonstrates the convergence behaviors, accuracy, robustness, and efficiency of the proposed method and discusses its advantages compared with other similar methods based on the same strategy. The convergence behaviors toward true solutions become straightforward upon increasing the order of the Gauss-Legendre quadrature rule $N_q$, and convergence is achieved with relatively small numbers of quadrature points. For various electrode materials, computational times are reduced compared with the method involving the inversion of the Hamiltonian matrix and a complex matrix generalized eigensolver such as ZGGEV. Sec. \ref{sec:sec5} demonstrates the validity of the proposed method through transmission calculations for Au atomic chain with a CO molecule. In both cases, the transmission spectra show good agreement with previous plane-wave DFT calculations; slight differences that might arise from the inconsistency of the type of pseudopotentials and exchange correlation functionals are also observed. Overall, the numerical tests validate the applicability of the proposed method for first-principles electron transport calculations. In Sec.~\ref{sec:Application}, we have studied the transport through silicene with the line defect as a function of the incident electron energy.

\section*{Acknowledgment}
S.~I.~ would like to express their gratitude to K. Hirose for fruitful discussions on the WFM method. This research was partially supported by MEXT as a social and scientific priority issue (Creation of new functional devices and high-performance materials to support next-generation industries) to be tackled by using post-K computer, and a Grant-in-Aid for JSPS
221 Research Fellow (Grant No. 16J00911). This work was supported in part by JST/CREST, JST/ACT-I (Grant No. JP-MJPR16U6), and MEXT KAKENHI (Grant No. 17K12690). S.~T.~ gratefully acknowledge the financial support from DFG through the Collaborative Research Center SFB 1238 (Project C01).

\appendix

\section{Shifted BiCG method and seed switching technique}
\label{appendix:B}

The BiCG method is one of the Krylov subspace methods for solving dual linear systems such as
\begin{equation}
Ax=b, \quad A^{\dagger}\tilde x=b, \label{eqn:Eq}\\
\end{equation}
where the matrix $A$ need not to be a Hermitian matrix. The algorithm updates the solution vectors $x$ and $\tilde x$ using the vectors $p$, $r$, $\tilde p$, and $\tilde r$ and the scalars $\alpha$ and $\beta$ via the following recurrences:
\begin{eqnarray}
\alpha_n &=& \frac{(\tilde r_n, r_n )}{(\tilde p_n, Ap_n)}, \\ 
x_{n+1} &=& x_{n}+\alpha_{n} p_{n},   \\ 
\tilde x_{n+1} &=& \tilde x_{n}+\bar \alpha_{n} \tilde p_{n},   \\ 
r_{n+1} &=& r_{n}-\alpha_{n} A p_{n},   \\ 
\tilde r_{n+1} &=& \tilde r_{n}-\bar \alpha_{n} A^{\dagger} \tilde p_{n},   \\ 
\beta_{n}&=&\frac{(\tilde r_{n+1},r_{n+1})}{(\tilde r_{n},r_{n})}, \\
p_{n+1}&=&r_{n+1}+\beta_{n}p_{n}, \\
\tilde p_{n+1}&=& \tilde r_{n+1}+\bar \beta_{n} \tilde p_{n},
\end{eqnarray}
where the initial conditions are set as $x_0=\tilde x_0=0$ and $r_0 = p_0 = \tilde r_0 = \tilde p_0 = b$.

Now, we focus on solving the $m$ sets of shifted dual linear systems:
\begin{equation}
(A+\sigma_i I)x^{(i)}=b, \quad (A^{\dagger}+\sigma_i I)\tilde x^{(i)}=b,
\label{eqn:dualshift}
\end{equation}
for $i=1,2,\ldots,m$, using the reference system $Ax=b$ and $A^{\dagger}\tilde x=b$, where $\sigma_i$ is a real-valued scalar shift and $I$ is the identity matrix. When we choose the initial conditions as $x^{(i)}_0=\tilde x^{(i)}_0=0$, the Krylov subspace of the reference system and shifted dual linear systems are identical. Consequently, the residual vectors $r_n^{(i)}$ and $\tilde r_n^{(i)}$ are collinear with $r_n$ and $\tilde r_n$, respectively, that is,
\begin{equation}
r_n^{(i)}=\frac{1}{\pi_n^{(i)}}r_n, \quad \tilde r_n^{(i)}=\frac{1}{\bar \pi_n^{(i)}} \tilde r_n,
\label{eqn:collinear}
\end{equation}
where $\pi_n^{(i)}$ is a scalar that is updated by the following recurrence:
\begin{equation}
\pi_{n+1}^{(i)}=(1+\frac{\beta_{n-1}\alpha_{n}}{\alpha_{n-1}}+\alpha_{n}\sigma_i)\pi_{n}^{(i)}-\frac{\beta_{n-1}\alpha_{n}}{\alpha_{n-1}}\pi_{n-1}^{(i)}.
\label{eqn:pi}
\end{equation}
Here, $\pi^{(i)}_0=\pi^{(i)}_{-1}=1$. By using the collinear relation given in Eq.~\eqref{eqn:collinear}, the shifted dual linear systems are updated by the following recurrences:
\begin{eqnarray}
\alpha_{n}^{(i)}&=&\frac{\pi_{n-1}^{(i)}}{\pi_{n}^{(i)}}\alpha_{n}, \\
\beta_{n}^{(i)}&=&\Bigl(\frac{\pi_{n-1}^{(i)}}{\pi_{n}^{(i)}}\Bigr)^2\beta_{n}, \\
x^{(i)}_{n+1} &=& x^{(i)}_{n}+\alpha^{(i)}_{n} p^{(i)}_{n},   \\ 
\tilde x^{(i)}_{n+1} &=& \tilde x^{(i)}_{n}+\bar \alpha^{(i)}_{n} \tilde p^{(i)}_{n},   \\ 
p^{(i)}_{n+1}&=&r^{(i)}_{n+1}+\beta^{(i)}_{n}p^{(i)}_{n}, \\
\tilde p^{(i)}_{n+1}&=& \tilde r^{(i)}_{n+1}+\bar \beta^{(i)}_{n} \tilde p^{(i)}_{n}. \label{eqn:endcollinear}
\end{eqnarray}
Because the recurrences in Eqs.~\eqref{eqn:collinear}-\eqref{eqn:endcollinear} consist of only scalar-scalar and scalar-vector products, the shifted dual linear systems can be solved very quickly, rather than applying the standard BiCG method to them.

The iterations continue until the residual norms of the entire system become sufficiently small. However, when the residual norm of the reference system becomes too small, the numerical precision of the residual vectors of shifted dual linear systems decreases. To avoid this problem, we use the seed switching technique that replaces the reference system with a shifted dual linear system whose residual norm is the largest in the entire system. To switch the reference system to the new one $\tilde{s}=$ arg max$ _{i\in I}$\{$||r^{i}_n||$\}, we need a scalar in Eq.~\eqref{eqn:pi} for the new reference system:
\begin{eqnarray}
\pi_{n}^{(\tilde{s},i)}&=&\frac{\pi_{n}^{(i)}}{\pi_{n}^{(\tilde{s})}}.
\label{eqn:seedswitch}
\end{eqnarray}
The seed switching technique for an arbitrary shift $\sigma$($\notin \{\sigma_1, \sigma_2,\ldots, \sigma_m\}$) is presented in Ref.~\onlinecite{Yamamoto2008}.

\section{Kronig-Penny model}
\begin{figure*}
\begin{center}
\includegraphics[width=120mm]{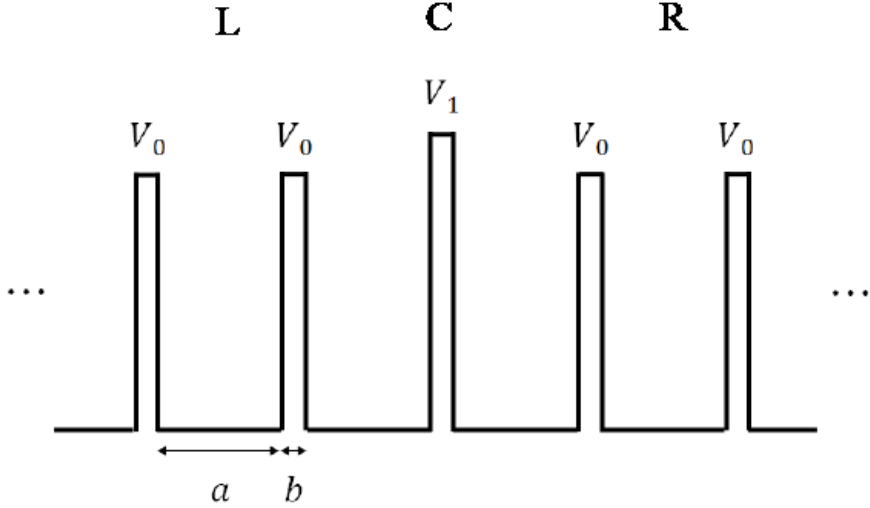}
\caption{An illustration of a one-dimensional system with square potential barriers. The parameters $a, b, V_0$, and $V_1$ represent the width of depths, width of barriers, barrier height in L and R regions, and barrier height in C region, respectively.}
\label{fig:kp_model}
\end{center}
\end{figure*}

\begin{figure*}
\begin{center}
\includegraphics[width=120mm]{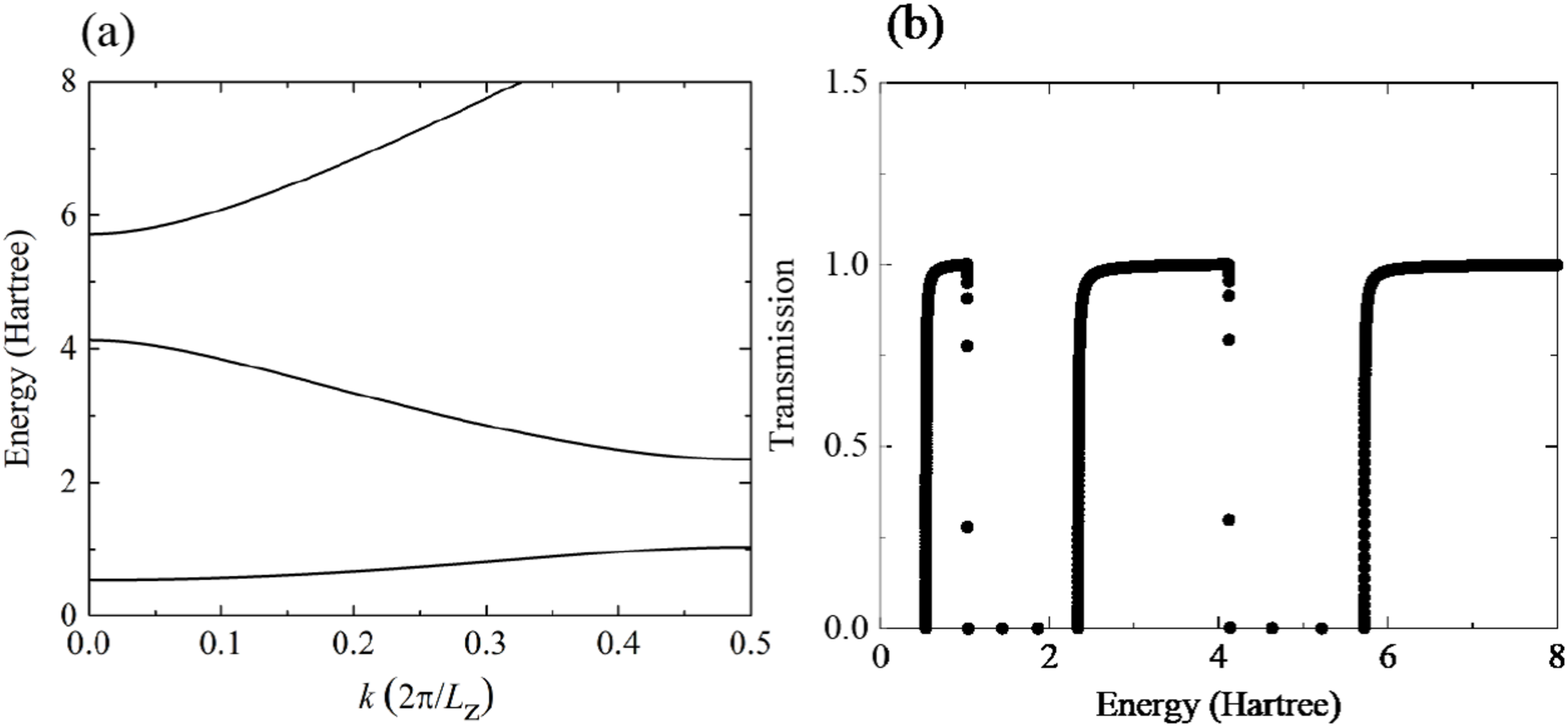}
\caption{(a) Energy dispersion of the Kronig-Penny model with periodic square potentials and (b) transmission spectra. The parameters in atomic units are set as $a = 2.0, b = 0.2, V_0=10$, and  $V_1=11$.}
\label{fig:kp_transmission}
\end{center}
\end{figure*}

We here discuss the effect of the small perturbation of the potential to the electron scattering in a one-dimensional system with square potentials. We divide the system into L, R, and C. We consider that L and R are semi-infinite electrodes with periodic square potentials and the barrier height of square potential in C is shifted. The parameters are given in Fig.~\ref{fig:kp_model}. Figs.~\ref{fig:kp_transmission}(a) and (b) show the energy dispersion for the periodic square potentials and transmission spectra, respectively. Naturally, we observe the scatterings at the band edges where the group velocity becomes zero. Note that the same tendency can be seen when varying parameters.

\providecommand{\noopsort}[1]{}\providecommand{\singleletter}[1]{#1}%

\end{document}